\documentclass{format/ieeeaccess}
\usepackage{amsmath,amssymb,amsbsy}
\usepackage{algorithm,algorithmicx,algpseudocode}
\usepackage{graphicx}
\graphicspath{{./figure/}}
\usepackage{arydshln,booktabs,multirow}
\usepackage{setspace,relsize}
\usepackage{hyperref}
\usepackage{balance}
\def\BibTeX{{\rm B\kern-.05em{\sc i\kern-.025em b}\kern-.08em
		T\kern-.1667em\lower.7ex\hbox{E}\kern-.125emX}}

\usepackage{caption,subcaption}
\usepackage[para]{threeparttable}

\newcommand{\eg}{\textit{e.g.}}
\newcommand{\ie}{\textit{i.e.}}

\newcommand{\plus}{PLUS}
\newcommand{\ptfm}{PLUS-TFM}
\newcommand{\prnn}{PLUS-RNN}
\newcommand{\plusb}{PLUS-RNN\textsubscript{BASE}}
\newcommand{\plusl}{PLUS-RNN\textsubscript{LARGE}}

\begin{document}
\history{Date of publication xxxx 00, 0000, date of current version xxxx 00, 0000.}
\doi{10.1109/ACCESS.2017.DOI}

\title{Pre-training of Deep Bidirectional Protein Sequence Representations with Structural Information}
\markboth
{Min \headeretal: Pre-training of Deep Bidirectional Protein Sequence Representations with Structural Information}
{Min \headeretal: Pre-training of Deep Bidirectional Protein Sequence Representations with Structural Information}


\author{
    \uppercase{Seonwoo Min}\authorrefmark{1,2},
    \uppercase{Seunghyun Park}\authorrefmark{3},
    \uppercase{Siwon Kim}\authorrefmark{1},
    \uppercase{Hyun-Soo Choi}\authorrefmark{4},
    \uppercase{Byunghan Lee}\authorrefmark{5} \IEEEmembership{Member, IEEE},
    and
    \uppercase{Sungroh Yoon}\authorrefmark{1,6} \IEEEmembership{Senior Member, IEEE}
}

\address[1]{Department of Electrical and Computer engineering, Seoul National University, Seoul 08826, South Korea}
\address[2]{LG AI Research, Seoul 07796, South Korea}
\address[3]{Clova AI Research, NAVER Corp., Seongnam 13561, South Korea}
\address[4]{Department of Computer Science and Engineering, Kangwon National University, Chuncheon 24341, South Korea}
\address[5]{Department of Electronic and IT Media Engineering, Seoul National University of Science and Technology, Seoul 01811, South Korea}
\address[6]{Interdisciplinary Program in Artificial Intelligence, ASRI, INMC, and Institute of Engineering Research, Seoul National University, Seoul 08826, South Korea}

\corresp{Corresponding author: Byunghan Lee (bhlee@seoultech.ac.kr) or Sungroh Yoon (sryoon@snu.ac.kr)}

\tfootnote{
This research was supported by the National Research Foundation (NRF) of Korea grants funded by the Ministry of Science and ICT (2018R1A2B3001628 (S.Y.), 2014M3C9A3063541 (S.Y.), 2019R1G1A1003253 (B.L.)), the Ministry of Agriculture, Food and Rural Affairs (918013-4 (S.Y.)), and the Brain Korea 21 Plus Project in 2021 (S.Y.). The funders had no role in study design, data collection and analysis, decision to publish, or preparation of the manuscript.
}
\begin{abstract}
Bridging the exponentially growing gap between the numbers of unlabeled and labeled protein sequences, several studies adopted semi-supervised learning for protein sequence modeling. In these studies, models were pre-trained with a substantial amount of unlabeled data, and the representations were transferred to various downstream tasks. Most pre-training methods solely rely on language modeling and often exhibit limited performance. In this paper, we introduce a novel pre-training scheme called \textbf{\plus}, which stands for \textbf{P}rotein sequence representations \textbf{L}earned \textbf{U}sing \textbf{S}tructural information. \plus\ consists of masked language modeling and a complementary protein-specific pre-training task, namely same-family prediction. \plus\ can be used to pre-train various model architectures. In this work, we use \plus\ to pre-train a bidirectional recurrent neural network and refer to the resulting model as \prnn.  Our experiment results demonstrate that \prnn\ outperforms other models of similar size solely pre-trained with the language modeling in six out of seven widely used protein biology tasks. Furthermore, we present the results from our qualitative interpretation analyses to illustrate the strengths of \prnn. \plus\ provides a novel way to exploit evolutionary relationships among unlabeled proteins and is broadly applicable across a variety of protein biology tasks. We expect that the gap between the numbers of unlabeled and labeled proteins will continue to grow exponentially, and the proposed pre-training method will play a larger role. All the data and codes used in this study are available at \href{https://github.com/mswzeus/PLUS}{https://github.com/mswzeus/PLUS}.
\vspace{-0.3cm}
\end{abstract}


\maketitle

\section{Introduction}

Proteins consisting of linear chains of amino acids are among the most versatile molecules in living organisms. They serve vital functions in biological mechanisms, \eg, transporting other molecules and providing immune protection \cite{berg2006}. The versatility of proteins is generally attributed to their diverse structures. Proteins naturally fold into three-dimensional structures determined by their amino acid sequences. These structures have a direct impact on their functions. 

With the development of next-generation sequencing technologies, protein sequences have become relatively more accessible. However, annotating a sequence with meaningful attributes is still time-consuming and resource-intensive. To bridge the exponentially growing gap between the numbers of unlabeled and labeled protein sequences, various {\it in silico} approaches have been widely adopted for predicting the characteristics of protein sequences \cite{holm1996}.

Sequence alignment is a key technique in computational protein biology. Alignment-based methods are used to compare protein sequences using carefully designed scoring matrices or hidden Markov models (HMMs) \cite{eddy2004, soding2005}. Correct alignments can group similar sequences, provide information on conserved regions, and help investigate uncharacterized proteins. However, its computational complexity increases exponentially with the number of proteins, and it has difficulties in identifying distantly related proteins. Homologous proteins sharing a common evolutionary ancestor can have high sequence-level variations \cite{creighton1993}. Therefore, a simple comparison of sequence similarities often fails to capture the global structural and functional similarities of proteins.

Building upon the success of deep learning, several studies proposed deep learning algorithms for computational protein biology. Some of these algorithms only use raw protein sequences, whereas others may use additional features \cite{min2017}. They have advanced the state-of-the-art (SOTA) for various protein biology tasks. However, development of these algorithms requires highly task-specific processes, \eg, training a randomly initialized model from scratch. It demands careful consideration of the model architectures and hyperparameters tailored for each task. Additional features, such as alignment-based features or known structural traits, may also be required for some tasks \cite{rao2019}. 

Semi-supervised learning, which leverages both unlabeled and labeled data, has been a long-standing goal of the machine learning community \cite{chapelle2009}. A semi-supervised learning algorithm pre-trains a universal model with a substantial amount of unlabeled data. Subsequently, it transfers the learned representations and fine-tunes the model with a small amount of labeled data for each downstream task. Now, the natural question is: can protein biology also take advantage of semi-supervised learning? According to the linguistic hypothesis \cite{alquraishi2019end}, naturally occurring proteins are not purely random. Evolutionary pressure constrains them to a learnable manifold where indispensable structures and functions are maintained. Thus, by observing many unlabeled protein sequences, we can obtain an implicit understanding of the language of proteins. Several studies have recently proposed pre-training methods for protein sequence representations \cite{yang2018, alley2019, bepler2019,  rives2019biological, strodthoff2019, heinzinger2019modeling, lu2020self, elnaggar2020prottrans}. They pre-trained models with language modeling (LM) and showed that pre-training helps in downstream protein biology tasks. However, according to the recent benchmark results from tasks assessing protein embeddings (TAPE) \cite{rao2019}, the current pre-training methods are often outperformed by task-specific models. This may be because most pre-training methods solely rely on LM to learn from unlabeled protein sequences. Therefore, a complementary protein-specific task for pre-training might be necessary to better capture the information contained within unlabeled protein sequences.

\begin{figure*}[t]
    \centering
	\includegraphics[width=0.9\textwidth]{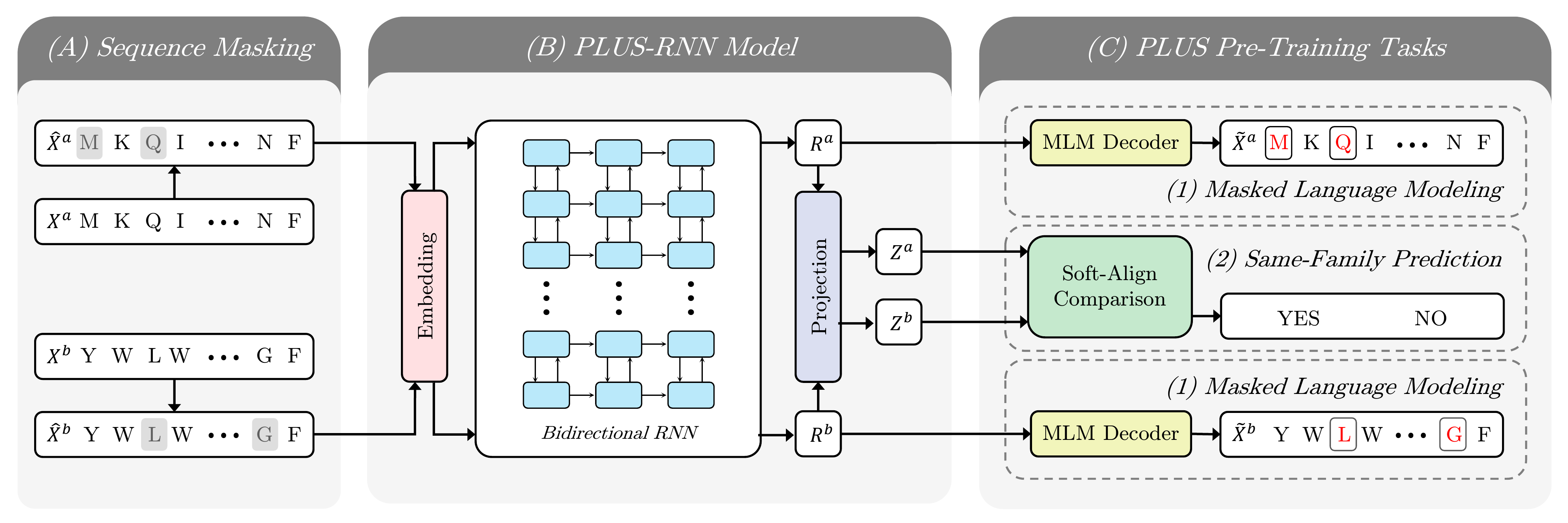}
  	\caption{Overview of \plus\ pre-training scheme (see Section III for details). (A) We randomly mask 15\% of amino acids (gray boxes) in each protein sequence $X^{a}$ and $X^{b}$. (B) \prnn\ transforms masked protein sequences $\hat{X}^{a}$ and $\hat{X}^{b}$ into sequences of bidirectional representations $R^{a}$ and $R^{b}$, respectively. (C) \plus\ consists of two pre-training tasks. Masked language modeling trains a model to predict the masked amino acids (colored red within white boxes of $\tilde{X}^{a}$ and $\tilde{X}^{b}$) given their contexts $\hat{X}^{a}$ and $\hat{X}^{b}$. Same-family prediction (SFP) trains a model to predict whether a pair of proteins belongs to the same protein family where soft-align comparison computes similarity score between projected sequences $Z^{a}$ and $Z^{b}$ of $R^{a}$ and $R^{b}$, respectively.}
    \label{fig:model_overview}
\end{figure*}

In this paper, we introduce a novel pre-training scheme for protein sequence modeling and name it \textbf{\plus}, which stands for \textbf{P}rotein sequence representations \textbf{L}earned \textbf{U}sing \textbf{S}tructural information. \plus\ consists of masked language modeling (MLM) and an additional complementary protein-specific pre-training task, same-family prediction (SFP). SFP leverages computationally clustered protein families \cite{finn2014} and helps to better capture the global structural information within unlabeled protein sequences. We use \plus\ to pre-train a bidirectional recurrent neural network (BiRNN) and refer to the resulting model as \prnn. Subsequently, this pre-trained universal model is fine-tuned on various downstream tasks without training randomly initialized task-specific models from scratch. Our experiment results demonstrate that \prnn\ outperforms other models of similar size solely pre-trained with the conventional LM in six out of seven widely used protein biology tasks. The seven tasks include three protein-level classification, two protein-level regression, and two amino-acid-level classification.  \plus\ provides a novel way to exploit evolutionary relationships among unlabeled proteins and is broadly applicable across a variety of protein biology tasks. Finally, we present the results from our qualitative interpretation analyses to illustrate the strengths of \prnn.

In summary, the contributions of our paper are as follows:
\begin{itemize}
    \item We introduce \plus, a novel pre-training scheme for bidirectional protein sequence representations.
    \item Consisting of MLM and protein-specific SFP pre-training tasks, \plus\ can better capture structural information contained within proteins. 
    \item \plus\ outperforms other models of similar sizes (solely pre-trained with the conventional LM) in six out of seven widely used protein biology tasks.
    \item We present qualitative interpretation analyses to better understand the strengths of our \plus\ framework.
\end{itemize}
\vspace{0.2cm}

\section{Related Works}

\subsection{Pre-training natural language representations}

Pre-training natural language representations has been the basis of natural language processing (NLP) research for a long time.  They use language modeling (LM) for pre-training natural language representations. The key idea is that ideal representations must convey syntactic and semantic information, and thus we must be able to use a representation of a token to predict its neighboring tokens. For example, embeddings from language models (ELMo) learned contextualized representations by adopting forward and reverse RNNs \cite{peters2018}. Given a sequence of tokens without additional labels, the forward RNN sequentially processes the sequence left-to-right. It is trained to predict the next token, given its history. The reverse RNN is similar but processes the sequence in reverse, right-to-left. After the pre-training, the hidden states of both RNNs are merged into a single vector representation for each token. Thus, the same token can be transformed into different representations based on its context.

The major limitation of ELMo is that RNNs are trained using unidirectional LM and simply combined afterward. As valuable information often comes from both directions, unidirectional LM is inevitably suboptimal. To address this problem, bidirectional encoder representations from Transformers (BERT) was proposed to pre-train bidirectional natural language representations using the Transformer model \cite{devlin2018}. Instead of the conventional LM, BERT utilizes an masked language modeling (MLM) pre-training task. It masks some input tokens at random and trains the model to predict them from the context. In addition, BERT includes a complementary NLP-specific pre-training task, next sentence prediction, which enables the learning of sentence relationships by training a model to predict whether a given pair of sentences is consecutive.

\vspace{0.1cm}

\subsection{Pre-training protein sequence representations}

NLP-based methods are historically adapted to learn protein sequence representations \cite{asgari2015, yang2018}. The previous methods most closely related to our paper are P-ELMo \cite{bepler2019} and UniRep \cite{alley2019}, which learn contextualized protein representations. P-ELMo is based on a two-phase algorithm. First, it trained forward and reverse RNNs using LM with an unlabeled dataset. hen, it adopted another bidirectional recurrent neural network (BiRNN) and further trained the model with an additional small labeled dataset. Note that the latter supervised training deviates from the goal of pre-training, namely, utilizing low human effort and large unlabeled datasets. UniRep used a unidirectional RNN with multiplicative long short-term memory hidden units \cite{krause2016}. Similarly, UniRep trained its model using conventional LM. 

Most methods have some common limitations and still often lag behind task-specific models \cite{rao2019}. First, some of them learn unidirectional representations from unlabeled datasets. Unidirectional representations are sub-optimal for numerous protein biology tasks, where it is crucial to assimilate global information from both directions. Note that we do not consider combination of two unidirectional representations as bidirectional representations since they were simply combined after the unidirectional pre-training. Second, most pre-training methods solely rely on LM to learn from unlabeled protein sequences. Although LM is a simple and effective task, a complementary pre-training task tailored for each data modality has been often the key to further improve the quality of representations in other domains. For instance, in NLP, BERT adopted the next sentence prediction task. In another example, ALBERT devised a complementary sentence order prediction task to model the inter-sentence coherence and yielded consistent performance improvements for downstream tasks \cite{lan2019}. Similarly, a complementary protein-specific task for pre-training might be necessary to better capture the information contained within unlabeled proteins.

\section{Methods}

We introduce \plus\ (Figure \ref{fig:model_overview}), a novel pre-training scheme for protein sequence modeling. Consisting of MLM and complementary protein-specific SFP pre-training tasks, PLUS can help a model to learn structurally contextualized bidirectional representations. In the following, we will explain the details of the pre-training procedures, fine-tuning procedures, and the model architecture.

\subsection{Pre-training procedure}
\plus\ can be used to pre-train various model architectures that transform a protein sequence $X = [\textbf{x}_{1}, \cdots, \textbf{x}_{n}]$, which has a variable-length $n$, into a sequence of bidirectional representations  $Z = [\textbf{z}_{1}, \cdots, \textbf{z}_{n}]$ with the same length. In this work, we use \plus\ to pre-train a BiRNN and refer to the resulting model as \prnn. The complete pre-training loss is defined as: 
\begin{equation*}
\mathcal{L}_{\mathrm{PT}} = \lambda_{\mathrm{PT}}\mathcal{L}_{\mathrm{MLM}} + (1 - \lambda_{\mathrm{PT}})\mathcal{L}_{\mathrm{SFP}}
\end{equation*}
where $\mathcal{L}_{\mathrm{MLM}}$ and $\mathcal{L}_{\mathrm{SFP}}$ are the MLM and SFP losses, respectively. We use $\lambda_{\mathrm{PT}}$ to control their relative importance (Appendix \ref{sec:appendix_training}).

\subsubsection{Task \#1: Masked Language Modeling (MLM)}
Given a protein sequence $X$, we randomly select 15\% of the amino acids. Then, for each selected amino acid $\textbf{x}_{i}$, we randomly perform one of the following procedures. For 80\% of the time, we replace $\textbf{x}_{i}$ with the token denoting an unspecified amino acid. For 10\% of the time, we randomly replace $\textbf{x}_{i}$ with one of the 20 amino acids. Finally, for the remaining 10\%, we keep $\textbf{x}_{i}$ intact. This is to bias the learning toward the true amino acids. For the probabilities of masking actions, we follow those used in BERT \cite{devlin2018}.

\prnn\ transforms a masked protein sequence $\hat{X}$ into a sequence of representations. Then, we use an MLM decoder to compute log probabilities from the representations for $\widetilde{X}$ over 20 amino acid types. The MLM task trains the model to maximize the probabilities corresponding to the masked ones. As the model is designed to accurately predict randomly masked amino acids given their contexts, the learned representations must convey syntactic and semantic information within proteins.

\subsubsection{Task \#2: Same-Family Prediction (SFP)}
Considering that additional pre-training tasks has been often the key for improving the quality of representations in other domains \cite{devlin2018, lan2019}, we devise a complementary protein-specific pre-training task. The SFP task trains a model to predict whether a given protein pair belongs to the same protein family. The protein family labels provide weak structural information and help the model learn structurally contextualized representations. Note that \plus\ is still a semi-supervised learning method; it is supervised by computationally clustered weak labels rather than human-annotated labels.

We randomly sample two protein sequences $X^a$ and $X^b$, from the training dataset. In 50\% of the cases, two sequences are sampled from the same protein family. For the other 50\%, they are randomly sampled from different families. \prnn\ transforms the protein pair into sequences of representations $Z^a = [\textbf{z}^{a}_{1}, \cdots, \textbf{z}^{a}_{n_{1}}]$ and $Z^b = [\textbf{z}^{b}_{1}, \cdots, \textbf{z}^{b}_{n_{2}}]$. Then, we use a soft-align comparison \cite{bepler2019} to compute their similarity score, $\hat{c}$, as a negative weighted sum of $l_{1}$-distances between every $\textbf{z}^{a}_{i}$ and $\textbf{z}^{b}_{j}$ pair:
\begin{equation*}
\hat{c} = - \frac{1}{C} \sum_{i=1}^{n_{1}} \sum_{j=1}^{n_{2}} \omega_{ij} \left\| \textbf{z}^{a}_{i} - \textbf{z}^{b}_{j} \right\|_{1}, \quad C = \sum_{i=1}^{n_{1}} \sum_{j=1}^{n_{2}} \omega_{ij},
\end{equation*}
where weight $\omega_{ij}$ of each $l_{1}$-distance is computed as

\begin{equation*}
\begin{gathered}
\omega_{ij} = 1 - (1 - \alpha_{ij})(1 - \beta_{ij}),\\
\alpha_{ij} = \frac{\exp({-\left\| \textbf{z}^{a}_{i} - \textbf{z}^{b}_{j} \right\|_{1}})}{\sum_{k=1}^{n_{2}} \exp({-\left\| \textbf{z}^{a}_{i} - \textbf{z}^{b}_{k} \right\|_{1}})}, \\
\beta_{ij} = \frac{\exp({-\left\| \textbf{z}^{a}_{i} - \textbf{z}^{b}_{j} \right\|_{1}})}{\sum_{k=1}^{n_{1}} \exp({-\left\| \textbf{z}^{a}_{k} - \textbf{z}^{b}_{j} \right\|_{1}})}.
\end{gathered}
\end{equation*}
Intuitively, we can understand the soft-align comparison as computing an {\it expected alignment score}, where they are summed over all the possible alignments. We suppose that the smaller the distance between representations, the more likely it is that the pair of amino acids is aligned. Then, we can consider $\alpha_{ij}$ as the probability that $\textbf{z}^{a}_{i}$ is aligned to $\textbf{z}^{b}_{j}$, considering all the amino acids from $Z^b$ (and vice versa for $\beta_{ij}$). As a result, $\hat{c}$ is the expected alignment score over all possible alignments with probabilities $\omega_{ij}$. Note that the negative signs are applied for converting distances into scores. Therefore, a higher value of $\hat{c}$ indicates that the pair of protein sequences is structurally more similar. 

Given the similarity score, the SFP output layer computes the probability that the pair belongs to the same protein family. The SFP task trains \prnn\ to minimize the cross-entropy loss between the true label and the predicted probability. As the model is designed to produce higher similarity scores for proteins from the same families, learned representations must convey global structural information.
\vspace{0.1cm}

\subsection{Fine-tuning procedure}
The fine-tuning procedure of \prnn\ follows the conventional usage of BiRNN-based prediction models. For each downstream task, we add one hidden layer and one output layer on top of the pre-trained model. Then, all the parameters are fine-tuned using task-specific datasets. The complete fine-tuning loss is defined as: 

\begin{equation*}
\mathcal{L}_{\mathrm{FT}} = \lambda_{\mathrm{FT}}\mathcal{L}_{\mathrm{MLM}} + (1 - \lambda_{\mathrm{FT}})\mathcal{L}_{\mathrm{TASK}}
\end{equation*}
where $\mathcal{L}_{\mathrm{TASK}}$ is the task-specific loss. $\mathcal{L}_{\mathrm{MLM}}$ is the regularization loss. We use $\lambda_{\mathrm{FT}}$ to control their relative importance (Appendix \ref{sec:appendix_training}).

The model's architectural modifications for the three types of downstream tasks are as follows. For tasks involving a protein pair, we use the same computations used in the SFP pre-training task. Specifically, we replace only the SFP output layer with a new output layer. For single protein-level tasks, we adopt an additional attention layer to aggregate variable-length representations into a single vector \cite{bahdanau2014}. Then, the aggregated vector is fed into the hidden and output layers. For amino-acid-level tasks, representations of each amino acid are fed into the hidden and output layers. 
\vspace{0.1cm}

\subsection{Model architecture}
\plus\ can be used to pre-train any model architectures that transform a protein sequence into a sequence of bidirectional representations. In this work, we use \prnn\ because of its superior sequential modeling capability and lower computational complexity. Refer to Appendix \ref{sec:appendix_transformer} for more detailed explanations of the advantages of \prnn\ over an alternative Transformer-based model, called \ptfm.

\begin{table*}[t]
 	\caption{Summarized results on protein biology benchmark tasks}
  	\label{table:result_summary}
  	\footnotesize
    \begin{threeparttable}
    \begin{tabular*}{\textwidth}{l@{\extracolsep{\fill}}ccccccc}
        \toprule
            & \multicolumn{3}{c}{Protein-level Classification} & \multicolumn{2}{c}{Protein-level Regression} & \multicolumn{2}{c}{Amino-acid-level Classification} \\
        \cmidrule(lr){2-4} \cmidrule(lr){5-6} \cmidrule(lr){7-8}
            Method & Homology (acc) & Solubility (acc)  & Localization (acc) & Stability ($\rho$) & Fluorescence ($\rho$) & SecStr (acc) & Transmembrane (acc) \\  
        \midrule    
            \ptfm               & 0.96 & \textbf{0.72} & 0.69 & 0.76 & 0.63 & 0.59 & 0.82 \\
            \plusb              & 0.96 & 0.70 & 0.69 & \textbf{\underline{0.77}} & 0.67 & 0.61 & \textbf{\underline{0.89}} \\
            \plusl              & \textbf{\underline{0.97}} & 0.71 & \textbf{0.70} & \textbf{\underline{0.77}} & \textbf{\underline{0.68}} & \textbf{0.62} & \textbf{\underline{0.89}} \\
        \addlinespace[0.6ex] \cdashline{1-8} \addlinespace[0.6ex]
            LM Pre-trained   & 0.95 & 0.64 & 0.54 & 0.73 & \textbf{\underline{0.68}} & 0.61 & 0.78 \\
        \addlinespace[0.6ex] \cdashline{1-8} \addlinespace[0.6ex]
            Task-specific SOTA  & 0.93 & 0.77 & 0.78 & 0.73 & 0.67 & 0.72 & 0.80 \\
        \bottomrule
    \end{tabular*}
    \begin{tablenotes}
        For each task, the best pre-trained model is in \textbf{bold}. It is \textbf{\underline{bold and underlined}} if it is the best when including the task-specific SOTA.
    \end{tablenotes}
    \end{threeparttable}
    \vspace{0.2cm}
\end{table*}

In this section, we explain the architecture of \prnn. First, an input embedding layer, $\mathrm{EM}$, embeds each amino acid $\textbf{x}_{i}$ into a $d_{e}$-dimensional dense vector $\textbf{e}_{i}$:
\begin{equation*}
E = [\textbf{e}_{1}, \cdots, \textbf{e}_{n}], \quad  \textbf{e}_{i} = \mathrm{EM}(\textbf{x}_{i}).
\end{equation*}
Then, a BiRNN of $L$-layers computes representations as a function of the entire sequence. We use long short-term memory as the basic unit of the BiRNN \cite{hochreiter1997}. In each layer, the BiRNN computes $d_{h}$-dimensional forward and backward hidden states ($\overrightarrow{\textbf{h}^{l}_{i}}$ and $\overleftarrow{\textbf{h}^{l}_{i}}$) and combines them into a hidden state $\textbf{h}^{l}_{i}$ using a non-linear transformation:
\begin{equation*}
\begin{gathered}
\overrightarrow{\textbf{h}^{l}_{i}} = \sigma(\overrightarrow{\textbf{W}^{l}_{x}}\textbf{h}^{l-1}_{i} + \overrightarrow{\textbf{W}^{l}_{h}}\textbf{h}^{l}_{i-1} + \overrightarrow{\textbf{b}^{l}}), \\
\overleftarrow{\textbf{h}^{l}_{i}} = \sigma(\overleftarrow{\textbf{W}^{l}_{x}}\textbf{h}^{l-1}_{i} + \overleftarrow{\textbf{W}^{l}_{h}}\textbf{h}^{l}_{i+1} + \overleftarrow{\textbf{b}^{l}}), \\ 
\textbf{h}^{l}_{i} = \sigma(\textbf{W}^{l}_{h}[\overrightarrow{\textbf{h}^{l}_{i}}; \overleftarrow{\textbf{h}^{l}_{i}}] + \textbf{b}^{l}) \quad for \quad l = 1, \cdots, L,
\end{gathered}
\end{equation*}
where $\textbf{h}^{0}_{i} = \textbf{e}_i$; $\textbf{W}$ and $\textbf{b}$ are the weight and bias vectors, respectively. We use the final hidden states $\textbf{h}^{L}_{i}$ as representations $\textbf{r}_i$ of each amino acid: 

\begin{equation*}
R = [\textbf{r}_{1}, \cdots, \textbf{r}_{n}], \quad  \textbf{r}_{i} = \textbf{h}^{L}_{i}.
\end{equation*}
We adopt an additional projection layer to obtain smaller $d_{z}$-dimensional representations $\textbf{z}_i$ of each amino acid with a linear transformation:
\begin{equation*}
Z = [\textbf{z}_{1}, \cdots, \textbf{z}_{n}], \quad  \textbf{z}_{i} = \mathrm{Proj}(\textbf{r}_{i}).
\end{equation*}
During pre-training, to reduce computational complexity, we use $R$ and $Z$ for the MLM and SFP tasks, respectively. During fine-tuning, we can use either $R$ or $Z$, considering the performance on development sets or based on the computational constraints.

We use two models with the fixed $d_{e}$ of 21 and $d_{z}$ of 100:
\vspace{0.1cm}
\begin{itemize}
    \item \plusb: $L$ = 3, $d_{h}$= 512, 15M parameters
    \item \plusl: $L$ = 3, $d_{h}$= 1024, 59M parameters
\end{itemize}
\vspace{0.1cm}
The hyperparameters (\ie, $L$ and $d_{h}$) of \plusb\ are chosen to match the BiRNN model architecture used in P-ELMo \cite{bepler2019}. However, as P-ELMo uses additional RNNs, \plusb\ has less than half the number of parameters that P-ELMo has (32M).
\vspace{0.2cm}

\section{Experiments} 

\subsection{Pre-training dataset}
We used Pfam (release 27.0) as the pre-training dataset \cite{finn2014}. After pre-processing (Appendix \ref{sec:appendix_training}), it contained 14,670,860 sequences from 3,150 families. The Pfam dataset provides protein family labels which were computationally pre-constructed by comparing sequence similarity using multiple sequence alignments and HMMs. Owing to the loose connection between sequence and structure similarities, the family labels only provide weak structural information \cite{elofsson1999}. Note that we did not use any human-annotated labels. Therefore, pre-training does not result in biased evaluations in fine-tuning tasks. The pre-training results are provided in the Appendix \ref{sec:appendix_pre-training}.
\vspace{0.1cm}

\subsection{Fine-tuning tasks}
We evaluated \prnn\ on seven protein biology tasks. The datasets were curated and pre-processed by the cited studies. In the main manuscript, we provide concise task definitions and evaluation metrics. Please refer to Appendix \ref{sec:appendix_fine-tuning} for more details. 

\textbf{Homology} is a protein-level classification task \cite{fox2013}. The goal is to classify the structural similarity level of a protein pair into \textit{family}, \textit{superfamily}, \textit{fold}, \textit{class}, or \textit{none}. We report the accuracy of the predicted similarity level and the Spearman correlation, $\rho$, between the predicted similarity scores and the true similarity levels. Furthermore, we provide the average precision (AP) from prediction scores at each similarity level. 

\textbf{Solubility} is a protein-level classification task \cite{khurana2018}. The goal is to predict whether a protein is \textit{soluble} or \textit{insoluble}. We report the accuracy of this task. 

\textbf{Localization} is a protein-level classification task \cite{almagro2017}. The goal is to classify a protein into one of 10 subcellular locations. We report the accuracy of this task.

\textbf{Stability} is a protein-level regression task \cite{rocklin2017}. The goal is to predict a real-valued proxy for intrinsic stability. This task is from TAPE \cite{rao2019}, and we report the Spearman correlation, $\rho$.

\textbf{Fluorescence} is a protein-level regression task \cite{sarkisyan2016}. The goal is to predict the real-valued fluorescence intensities. This task is from TAPE, and we report the Spearman correlation, $\rho$.

\textbf{Secondary structure (SecStr)} is an amino-acid-level classification task \cite{klausen2019}. The goal is to classify each amino acid into eight or three classes, that describe its local structure. This task is from TAPE. We report both the three-way and eight-way classification accuracies (Q8/Q3) of this task.

\textbf{Transmembrane} is an amino-acid-level classification task \cite{tsirigos2015}. The goal is to detect amino acid segments that cross the cell membrane. We report the accuracy of this task.
\vspace{0.1cm}

\subsection{Baselines}
We provided several baselines for comparative evaluations. Note that since up-scaling of models and datasets often provide performance improvements, we only considered those with a similar scale of model sizes and pre-training datasets to focus on evaluating the pre-training schemes.

First, in all the tasks, we used two baselines: P-ELMo and \ptfm. The former has a model architecture similar to \plusb; thus, it can show the effectiveness of the pre-training scheme. The latter is pre-trained with \plus, so it can show the effectiveness of the BiRNN compared to the Transformer architecture.

Second, for the tasks from TAPE, we provide their reported baselines: P-ELMo, UniRep, TAPE-TFM, TAPE-RNN, and TAPE-ResNet. Note that these comparisons are in their favor, as they used a larger pre-training dataset (32M proteins from Pfam release 32.0). The TAPE baselines can demonstrate that \prnn\ outperforms models of similar size solely pre-trained with the LM.

Finally, we benchmarked \prnn\ against task-specific SOTA models trained from scratch. If no deep learning-based baseline exists for a given task, we provided RNN\textsubscript{BASE} and RNN\textsubscript{LARGE} models without pre-training. The comparison with those exploit additional features can help us identify the tasks for which the proposed pre-training scheme is most effective and help us understand its current limitations. 
\vspace{0.1cm}

\subsection{Summary of fine-tuning results}

Table \ref{table:result_summary} presents the summarized results for the benchmark tasks. Specifically, we show the best results from two categories: LM pre-trained models and task-specific SOTA models. Refer to Appendix \ref{sec:appendix_fine-tuning} for detailed fine-tuning results.

The \plusl\ model outperformed models of similar size solely pre-trained with the conventional LM in six out of seven tasks. Considering that some pre-trained models exhibited higher LM capabilities (Appendix \ref{sec:appendix_pre-training}), it can be speculated that the protein-specific SFP pre-training task contributed to the improvement. In the ablation studies, we further explained the relative importance of each aspect of \prnn\ (Appendix \ref{sec:appendix_ablation}). Although \ptfm\ had almost twice as many parameters as \plusl\ (110M vs. 59M), it exhibited inferior performance in most tasks. We infer that this is because it disregarded the \textit{locality bias} (Appendix \ref{sec:appendix_transformer}).

\begin{table}[t]
    \centering
 	\caption{Detailed Homology prediction results}
  	\label{table:results_homology}
  	\footnotesize
    \begin{threeparttable}
    \begin{tabular*}{\columnwidth}{@{\extracolsep{\fill}}lcccccc}
        \toprule
            &  \multicolumn{2}{c}{Overall} & \multicolumn{4}{c}{Per-level AP} \\
        \cmidrule(lr){2-3} \cmidrule(lr){4-7}
            Method & acc & $\rho$ & Class & Fold & Superfamily & Family \\
        \midrule    
            \ptfm   & 0.96 & \textbf{\underline{0.70}} &	0.94 &	0.91 &	0.95 &	\textbf{\underline{0.67}} \\
            \plusb  & 0.96 & 0.69 &	0.94 &	0.90 &	0.94 &	0.66 \\
            \plusl  & \textbf{\underline{0.97}} & \textbf{\underline{0.70}} &	\textbf{\underline{0.95}} &	\textbf{\underline{0.92}} &	\underline{\textbf{0.96}} &	0.66 \\
        \addlinespace[0.6ex] \cdashline{1-7} \addlinespace[0.6ex]
            P-ELMo    & 0.95 & 0.69 & 0.90 & 0.88 & 0.94 & 0.65  \\
            P-ELMo\tnote{\textdagger} & 0.95 & 0.69 & 0.91 & 0.90 & 0.95 & 0.65  \\
        \addlinespace[0.6ex] \cdashline{1-7} \addlinespace[0.6ex]
            NW-align\tnote{\textdagger}  & 0.78 &	0.22 &	0.31 &	0.41 &	0.58 &	0.53 \\
            HHalign\tnote{\textdagger}   & 0.79 &	0.23 &	0.40 &	0.62 &	0.86 &	0.52 \\
            TMalign\tnote{\textdagger}   & 0.81 &	0.37 &	0.55 &	0.85 &	0.83 &	0.57 \\
            RNN\textsubscript{BASE}    & 0.93 &	0.66 &	0.86 &	0.80 &	0.89 &	0.62 \\
            RNN\textsubscript{LARGE}   & 0.83 &	0.52 &	0.66 &	0.46 &	0.52 &	0.39 \\
        \bottomrule
    \end{tabular*}
    \begin{tablenotes}
            \item[\textdagger] Excerpted from P-ELMo.
    \end{tablenotes}
    \end{threeparttable}
    \vspace{0.3cm}
\end{table}
\begin{table}[t]
    \centering
 	\caption{Detailed SecStr prediction results}
  	\label{table:results_secstr}
  	\footnotesize
    \begin{threeparttable}
    \begin{tabular*}{\columnwidth}{l@{\extracolsep{\fill}}cccccc}
        \toprule
              & \multicolumn{2}{c}{CB513} & \multicolumn{2}{c}{CASP12} & \multicolumn{2}{c}{TS115} \\
        \cmidrule(){2-3} \cmidrule(){4-5} \cmidrule(){6-7}
            Method & Q8 & Q3 & Q8 & Q3 & Q8 & Q3 \\
        \midrule    
            \ptfm & 0.59 &	0.73 &	0.57 &	0.71 &	0.65 &	0.77  \\
            \plusb & 0.61 &	0.75 &	0.60 &	0.72 &	0.66 &	0.78  \\
            \plusl & \textbf{0.62} &	\textbf{0.77} &	\textbf{0.60} &	\textbf{0.73} &	\textbf{0.68} &  \textbf{0.79}  \\
        \addlinespace[0.6ex] \cdashline{1-7} \addlinespace[0.6ex]
            P-ELMo\tnote{\textasteriskcentered}    &  0.61 &	\textbf{0.77} &	0.54 &	0.68 &	0.63 &	0.76 \\
            P-ELMo\tnote{\textdagger} &  0.58 &	0.73 &	0.57 &	0.70 &	0.65 &	0.76 \\
            UniRep\tnote{\textdagger} &  0.57 &	0.73 &	0.59 &	0.72 &	0.63 &	0.77 \\
            TAPE-TFM\tnote{\textdagger} &  0.59 &	0.73 &	0.59 &	0.71 &	0.64 &	0.77 \\
            TAPE-RNN\tnote{\textdagger} &  0.59 &	0.75 &	0.57 &	0.70 &	0.66 &	0.78 \\
            TAPE-ResNet\tnote{\textdagger} &  0.58 &	0.75 &	0.58 &	0.72 &	0.64 &	0.78 \\
        \addlinespace[0.6ex] \cdashline{1-7} \addlinespace[0.6ex]
            NetSurfP-2.0\tnote{\textdaggerdbl} &  0.72 &	0.85 &	0.70 &	0.82 &	0.75 &	0.86 \\
        \bottomrule
    \end{tabular*}
    \begin{tablenotes}
            \item[\textdagger] Excerpted from TAPE.
            \item[\textdaggerdbl] Excerpted from \cite{klausen2019}.
    \end{tablenotes}
    \end{threeparttable}
    \vspace{0.3cm}
\end{table}
\begin{figure}[t]
    \centering
	\includegraphics[width=\columnwidth]{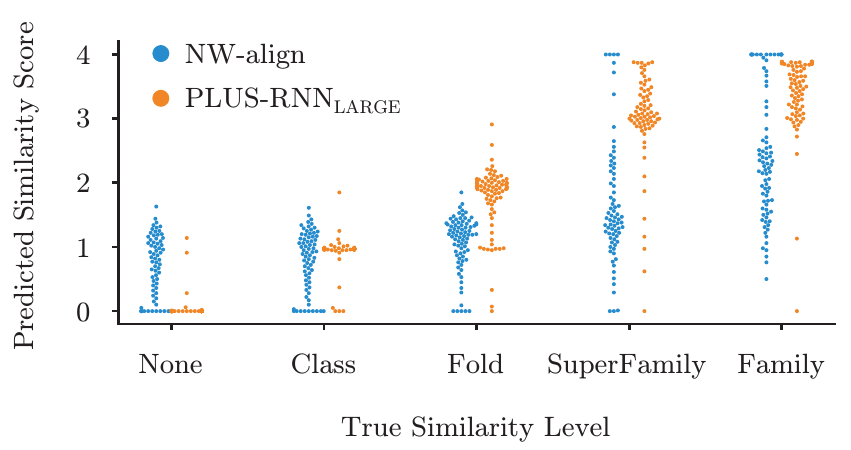}
   	\caption{Plot of predicted similarity scores and true similarity levels. Results from NW-align (left) and \plusl\ (right) are presented in each similarity level. Figure best viewed in color.}
    \label{fig:homology_plot}
\end{figure}
\begin{figure*}
    \centering
	\includegraphics[width=0.92\textwidth]{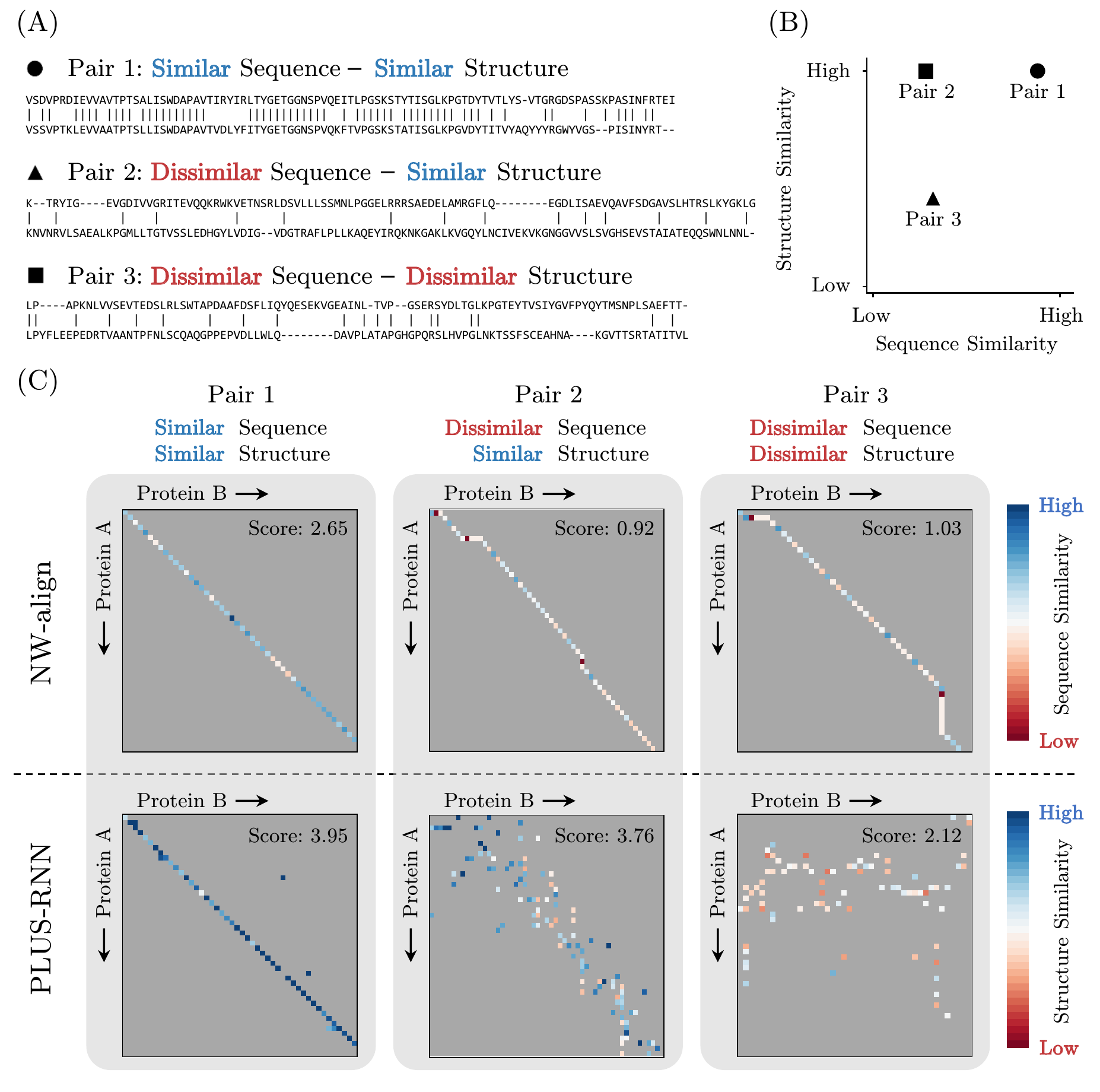}
   	\caption{Homology interpretation (A) We investigate three types of protein pairs: a similar sequence - similar structure pair, a dissimilar sequence - similar structure pair, and a dissimilar sequence - dissimilar structure pair. (B) The sequence and structure similarities of each pair are defined by NW-align scores and Homology dataset labels, respectively. (C) Heatmaps of NW-align of raw amino acids and the soft-alignment of \prnn\ representations for the three pairs. Owing to space limitations, we only show the top left quadrant of the heatmaps. Figure best viewed in color.}
    \label{fig:homology_interpretations}
\end{figure*}

We compared \plusl\ with task-specific SOTA models. Although the former performed better in some tasks, it still lagged behind on the others. The results indicated that tailored models with additional features provide powerful advantages that could not be learned through pre-training. A classic example is the use of position-specific scoring matrices generated from multiple sequence alignments. We conjectured that simultaneous observation of multiple proteins could facilitate evolutionary information. In contrast, current pre-training methods use millions of proteins; however, they still consider each one individually. The relatively small performance improvement from \plus\ could also be explained by the fact that the SFP task only utilizes pairwise information. We expect that investigating multiple proteins during pre-training might be the key to a superior performance over the task-specific SOTA models
\vspace{0.1cm}

\subsection{Detailed Homology and SecStr results}
We present detailed evaluation results for the Homology and SecStr tasks. We chose these two tasks because they are representative protein biology tasks relevant to global and local structures, respectively. Improved results on the former can lead to the discovery of new enzymes and antibiotic-resistant genes \cite{tavares2013}. The latter is important for understanding the function of proteins in cases where evolutionary structural information is not available \cite{klausen2019}.

The detailed Homology prediction results are listed in Table \ref{table:results_homology}. The results show that \plusl\ outperformed both P-ELMo and task-specific models. In contrast to RNN\textsubscript{LARGE}, which exhibited overfitting owing to the limited labeled training data, \plus\ pre-training enabled us to take advantage of the large model architecture. The correlation differences among \plusl\ (0.697), \plusb\ (0.693), and P-ELMo (0.685) were small; however, they were statistically significant with p-values lower than $10^{-15}$ \cite{steiger1980}. The per-level AP results helped us further examine the level of structural information captured by the pre-training. The largest performance improvement of \prnn\ comes at the higher \textit{class} level rather than the lower \textit{family} level. This indicates that even though Pfam family labels tend to be structurally correlated with the Homology task \textit{family} levels \cite{creighton1993}, they are not the decisive factors for performance improvement. Instead, \plus\ pre-training incorporates weak structural information and facilitates inferring higher-level global structure similarities. 

The detailed SecStr prediction results are listed in Table \ref{table:results_secstr}. CB513, CASP12, and TS115 denote the SecStr test datasets. The results show that \plusl\ outperformed all the other models of similar size pre-trained solely with LM. This demonstrates that the SFP task complements the LM task during pre-training and helps in learning improved structurally contextualized representations. However, \plusl\ still lagged behind task-specific SOTA models that employ alignment-based features. We infer that this limitation might be attributable to the following two factors. First, as previously stated, \plus\ only utilizes pairwise information, rather than simultaneously examining multiple proteins during pre-training. Second, the SFP task requires an understanding of the global structures, and thus the local structures are relatively negligible. Therefore, we believe that devising an additional pre-training task relevant to local structural information would improve the performance on the SecStr task.
\vspace{0.1cm}

\subsection{Qualitative analyses}
To better understand the strengths of \plus\ pre-training, we provide its qualitative analyses. We examined the Homology task to interpret how the learned protein representations help infer the global structural similarities of proteins

To compare two proteins, \prnn\ used soft-align to compute their similarity score, $\hat{c}$. Even though there was one more computation by the output layer for the Homology prediction output, we could use the similarity scores to interpret \prnn. Note that using the penultimate layer for model interpretation is a widely adopted approach in the machine learning community \cite{zintgraf2017}. 

Figure \ref{fig:homology_plot} shows a scatter plot of the predicted similarity scores and true similarity levels. For comparison, we also show the NW-align results based on the BLOSUM62 scoring matrix \cite{eddy2004}. The plot shows that NW-align often produces low similarity scores for protein pairs from the same \textit{family}. This is because of high sequence-level variations, which result in dissimilar sequences having similar structures. In contrast, \plusl\ produces high similarity scores for most protein pairs from the same \textit{family}. 

Furthermore, we examined three types of protein pairs: (1) a similar sequence-similar structure pair, (2) a dissimilar sequence-similar structure pair, and (3) a dissimilar sequence-dissimilar structure pair (Figure \ref{fig:homology_interpretations}(A) and (B)). Note that similar sequence-dissimilar structure pairs did not exist in the Homology datasets. The sequence and structure similarities were defined by NW-align scores and Homology dataset labels, respectively. The pairs with similar structures were chosen from the same \textit{family}, and those with dissimilar structures were chosen from the same \textit{fold}. Figure \ref{fig:homology_interpretations}(C) shows the heatmaps of the NW-align of raw amino acids and soft-alignment of \prnn\ representations ($\omega_{ij}$) for the three pairs. Owing to space limitations, we only show the top left quadrant of the heatmaps. Each cell in the heatmap indicates the corresponding amino acid pairs from proteins A and B. Blue denotes high sequence similarity in NW-align and high structure similarity in \prnn.

First, we compared the pairs having similar structures (the first and second columns in Figure \ref{fig:homology_interpretations}(C)). The heatmaps show that NW-align successfully aligned the similar-sequence pair, resulting in a score of 2.65. However, it failed for the dissimilar-sequence pair, with a score of 0.92. This supports the observation that comparing raw sequence similarities cannot identify the correct structural similarities. In contrast, the soft-alignment of \prnn\ representations was successful for both similar and dissimilar sequences, with scores of 3.95 and 3.76, respectively. Next, we compared the second and third pairs. Although only the second pair had similar structures, NW-align failed for both and even yielded a higher score of 1.03 for the third pair. In contrast, regardless of the sequence similarities, the soft-alignment of \prnn\ representations correctly decreased only for the third pair, with dissimilar structures having a score of 2.12. Therefore, the interpretation results confirmed that the learned representations from \prnn\ are structurally contextualized and perform better in inferring global structure similarities. 
\vspace{0.2cm}

\section{Conclusion}

In this work, we presented \plus, a novel pre-training scheme for bidirectional protein sequence representations. Consisting of the MLM and protein-specific SFP pre-training tasks, PLUS outperformed the conventional LM pre-training methods by capturing structural information contained within the proteins. \plus\ can be used to pre-train various model architectures. In this work, we used \prnn\ because of its superior sequential modeling capability and lower computational complexity. \prnn\ outperformed models of similar size solely pre-trained with the conventional LM in six out of seven protein biology tasks. To better understand its strengths, we also provided the results from our qualitative interpretation analyses.

We expect that the gap between the numbers of unlabeled and labeled proteins will continue to grow exponentially, and pre-training methods will play a larger role. We plan to extend this work in several directions. First, considering that \prnn\ is powerful for inferring global structural information, we are interested in a more refined prediction of protein structures \cite{kryshtafovych2019}. Second, although pre-training helps, our scheme still lags behind task-specific models in some tasks. We think that this limitation comes from weaknesses in learning evolutionary information. We believe that there is still considerable room for improvement. Investigation of multiple proteins during pre-training, as in the alignment, could be the key \cite{poplin2018}.

\appendices
\section{Details on Training Procedures}

\label{sec:appendix_training}
All models were implemented in PyTorch \cite{paszke2017}. We used the NAVER smart machine learning environment for pre-training \cite{sung2017}. In the following subsections, we explain the details of the pre-training and fine-tuning procedures.

\subsection{Pre-training procedure}
We used Pfam (release 27.0) as the pre-training dataset \cite{finn2014}. Moreover, we divided the training and test sets in a random 80\%/20\% split and filtered out sequences shorter than 20 amino acids. Additionally, for the training set, we removed families containing fewer than 1,000 proteins. This resulted in 14,670,860 sequences from 3,150 families being utilized for the pre-training of \plus. For the test dataset, we sampled 100,000 pairs from the test split.

We pre-trained \prnn\ with a batch size of 64 sequences for 900,000 steps, which is approximately four epochs over the training dataset. We used the Adam optimizer \cite{kingma2014} with a learning rate of 0.001, $\beta_{1} = 0.9$, $\beta_{2} = 0.999$, and without weight decay and dropout. Default $\lambda_{\mathrm{PT}}$ was 0.7.

For pre-training \ptfm, we used different filtering conditions due its computational complexity. The minimum and maximum lengths of a protein were set to 20 and 256, respectively. The minimum number of proteins for a family was set to 1000. This resulted in 11,956,227 sequences from 2,657 families. We pre-trained \ptfm\ with a batch size of 128 for 930,000 steps, which is approximately 10 epochs over the training dataset. We used the Adam optimizer with a learning rate of 0.0001, $\beta_{1} = 0.9$, $\beta_{2} = 0.999$, $L_{2}$ weight decay of 0.01, a linearly decaying learning rate with warmup over the first 10\% steps, and a dropout probability of 0.1. Default $\lambda_{\mathrm{PT}}$ was 0.7.

\subsection{Fine-tuning procedure}
When fine-tuning \prnn, most model hyperparameters were the same as those during the pre-training. The commonly used hyperparameters were as follows; we fine-tuned the \prnn\ with a batch size of 32 for 20 epochs. We used the Adam optimizer with a smaller learning rate of 0.0005, $\beta_{1} = 0.9$, $\beta_{2} = 0.999$, and without weight decay. For the other hyperparameters, we chose the configurations that performed best on the development sets for each task. The possible configurations were as follows:
\begin{itemize}
    \item Number of units in the added output layer: 128, 512
    \item Usage of the projection layer: True, False
    \item $\lambda_{\mathrm{FT}}$: 0, 0.3, 0.5
\end{itemize}

For fine-tuning \ptfm, we used the following hyperparameters: batch size of 32 for 20 epochs, the Adam optimizer with a smaller learning rate of 0.00005, $\beta_{1} = 0.9$, $\beta_{2} = 0.999$, and without weight decay. Default $\lambda_{\mathrm{FT}}$ was 0.3.

For fine-tuning both \prnn\ and \ptfm\ for the Homology task, we additionally followed the procedures of \cite{bepler2019}. In each epoch, we sampled 100,000 protein pairs using the smoothing rule: the probability of sampling a pair with similarity level $t$ is proportional to $N^{0.5}_{t}$, where $N_{t}$ is the number of protein pairs with similarity level $t$.

\section{Details on \ptfm}

\label{sec:appendix_transformer}

\subsection{Transformer architecture}
The key element of the Transformer is a self-attention layer composed of multiple attention heads \cite{vaswani2017}. Given an input sequence, $X = [\textbf{x}_{1}, \cdots, \textbf{x}_{n}]$, an attention head computes the output sequence, $Z = [\textbf{z}_{1}, \cdots, \textbf{z}_{n}]$. Each token is a weighted sum of values computed by a weight matrix $\textbf{W}^{V}$:

\begin{equation*}
\textbf{z}_{i} = \sum_{j=1}^{n} \alpha_{ij} (\textbf{x}_{j}\textbf{W}^{V}).
\end{equation*}
Each attention coefficient, $\alpha_{ij}$, is the output of a softmax function applied to the dot products of the query with all keys, which are computed using $\textbf{W}^{Q}$ and $\textbf{W}^{K}$:

\begin{equation*}
\label{equation:self_attention}
\alpha_{ij} = \frac{\exp(\textbf{e}_{ij})}{\sum_{k=1}^{n} \exp(\textbf{e}_{ik})}, \qquad
\textbf{e}_{ij} = \frac{(\textbf{x}_{i}\textbf{W}^{Q})(\textbf{x}_{j}\textbf{W}^{K})^{T}}{\sqrt{d_{z}}},
\end{equation*}
where $d_{z}$ is the output token dimension. The self-attention layer directly performs $\textit{O}(1)$ computations for all the pairwise tokens, whereas a recurrent layer requires $\textit{O}(n)$ sequential computations for the farthest pair. This allows easier traversal for forward and backward signals and thus better captures long-range dependencies.

The model architecture of \ptfm\ is analogous to the BERT\textsubscript{BASE} model, consisting of 110M parameters. Because of its significant computational burden, which scale quadratically with the input length, we pre-trained \ptfm\ using only protein pairs shorter than 512 amino acids, following the procedures used in BERT. 
\vspace{0.5cm}

\subsection{Advantages of \prnn\ over \ptfm}
\plus\ can be used to pre-train various model architectures including BiRNN and the Transformer. The resulting models are referred to as \prnn\ and \ptfm, respectively. In this work, we mainly used \prnn, because of its two advantages over \ptfm. First, it is more effective for learning the sequential nature of proteins. The self-attention layer of the Transformer performs dot products between all pairwise tokens regardless of their positions within the sequence. In other words, it provides an equal opportunity for local and long-range contexts to determine the representations. Although this facilitates the learning of long-range dependencies, the downside is that it completely ignores the \textit{locality bias} within a sequence. This is particularly problematic for protein biology, where local amino acid motifs often have significant structural and functional implications \cite{bailey2006}. In contrast, RNN sequentially processes a sequence, and local contexts are naturally more emphasized.

Second, \prnn\ provides lower computational complexity. Although the model hyperparameters have an effect, the Transformer-based models generally demand a larger number of parameters than RNNs \cite{vaswani2017}. Furthermore, the computations between all pairwise tokens in the self-attention layer impose a considerable computational burden, which scales quadratically with the input sequence length. Considering that pre-training typical Transformer-based models handling 512 tokens already requires tremendous resources \cite{devlin2018}, it is computationally difficult to use Transformers to manage longer protein sequences, even up to a few thousand amino acids.
\vspace{0.5cm}

\section{Pre-training Results}

\label{sec:appendix_pre-training}

\begin{figure}[t]
    \centering
	\includegraphics[width=0.8\columnwidth]{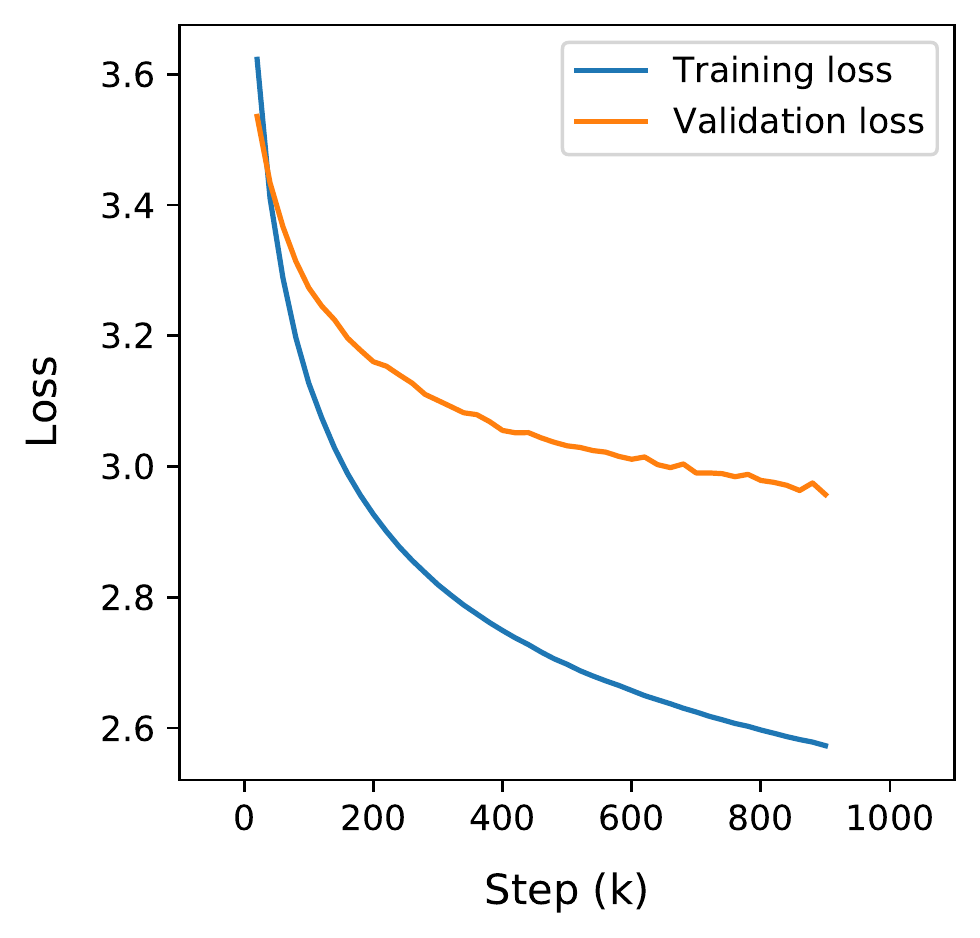}
   	\caption{Pre-training curve}
    \label{fig:pre-training_curve}
\end{figure}

Figure \ref{fig:pre-training_curve} shows the pre-training cure where training and validation losses are plotted concerning the parameter update steps. While the pre-training had not converged at 900,000 steps, we used the early stopping technique due to the time limitations. It required approximately three weeks to pre-train \plusl\ for 900,000 steps.

Table \ref{table:result_pre-training} lists the test accuracies for the MLM and SFP pre-training tasks. Only the models pre-trained with \plus\ were evaluated for the SFP task. Note that our experiments and TAPE used different test datasets; care should be taken in comparing them. Nonetheless, we can still indirectly compare them, considering the following. First, both the test datasets comprised randomly sampled proteins from different versions of the Pfam dataset (27.0 for \plus\ and 32.0 for TAPE). Second, P-ELMo was evaluated in both datasets and showed similar LM accuracies. This indicates that the difference between the two datasets is negligible.

We can see that some models have lower LM accuracies than others. However, the lower LM capability does not precisely correspond to the performance in fine-tuning tasks. This discrepancy has been previously observed in TAPE, and it can also be observed in the following sections. In terms of SFP, all the models pre-trained with \plus\ exhibited high accuracies. As the Pfam families were constructed based only on sequence similarities, a pair of analogous sequences would probably be from the same family. Despite its simplicity, we empirically demonstrated that the SFP complements the MLM by encouraging the models to compare protein representations during pre-training.

\begin{table}[t!]
    \centering
 	\caption{Results on pre-training tasks}
  	\label{table:result_pre-training}
  	\footnotesize
    \begin{threeparttable}
    \begin{tabular*}{\columnwidth}{l@{\extracolsep{\fill}}cc}
        \toprule
            Method & (M)LM (acc) & SFP (acc) \\
        \midrule    
            \ptfm & 0.37 & \textbf{0.98} \\ 
            \plusb & 0.33 & 0.96 \\ 
            \plusl & 0.37 & 0.97 \\ 
        \addlinespace[0.6ex] \cdashline{1-3} \addlinespace[0.6ex]    
            P-ELMo\tnote{\textasteriskcentered} & 0.29 & - \\
            P-ELMo\tnote{\textdagger} & 0.28 & -   \\
            UniRep\tnote{\textdagger} & 0.32 & -   \\     
            TAPE-TFM\tnote{\textdagger}    & \textbf{0.45} & -   \\
            TAPE-RNN\tnote{\textdagger}   & 0.40 & -   \\
            TAPE-ResNet\tnote{\textdagger} & 0.41 & -   \\
        \bottomrule
    \end{tabular*}
    \begin{tablenotes}
            \item[\textasteriskcentered] Our experiments (Pfam 27.0).
            \item[\textdagger] Excerpted from TAPE (Pfam 32.0).
    \end{tablenotes}
    \end{threeparttable}
\end{table}
\vspace{0.5cm}

\section{Fine-tuning Results}

\label{sec:appendix_fine-tuning}

\subsection{Homology}
For the Homology task, we used the SCOPe datasets \cite{fox2013}, which were pre-processed and provided by \cite{bepler2019}. The dataset was filtered with a maximum sequence identity of 95\% and split into 20,168 training, 2,240 development, and 5,602 test sequences. For the development and test datasets, we used the sampled 100,000 pairs from each dataset.

The detailed Homology prediction results are listed in Table 2. We report the excerpted baseline results from \cite{bepler2019}. NW-align computed the similarity between proteins based on sequence alignments with the BLOSUM62 substitution matrix, with a gap open penalty of -11 and a gap extension penalty of -1 \cite{eddy2004}. HHalign conducted multiple sequence alignments and searched similar sequences with HHbits \cite{soding2005, remmert2012}. TMalign performed structure alignments and scored a pair as the average of target-to-query and query-to-target scores \cite{zhang2005}. Note that we additionally normalized the scores from NW-align and HHalign with the sum of the lengths of protein pairs. This results in minor correlation improvements compared with the results reported by \cite{bepler2019}. 

\subsection{Solubility} 
For the Solubility task, we used the pepeDB datasets \cite{berman2009, chang2014}, which were pre-processed and provided by \cite{khurana2018}. The dataset was split into 62,478 training, 6,942 development, and 2,001 test sequences. The training dataset was filtered with a maximum sequence identity of 90\% to remove data redundancy. Furthermore, any sequences with more than 30\% sequence identity to those from the test dataset were removed to avoid any bias from homologous sequences.

\begin{table}[t!]
    \centering
 	\caption{Detailed solubility prediction results}
  	\label{table:results_solubility}
  	\footnotesize
    \begin{threeparttable}
    \begin{tabular*}{\columnwidth}{l@{\extracolsep{\fill}}c}
        \toprule
            Method & acc \\
        \midrule    
            \ptfm & \textbf{0.72}  \\
            \plusb & 0.70  \\
            \plusl & 0.71  \\
        \addlinespace[0.6ex] \cdashline{1-2} \addlinespace[0.6ex]
            P-ELMo &  0.64 \\
        \addlinespace[0.6ex] \cdashline{1-2} \addlinespace[0.6ex]
            SOLPro\tnote{\textdagger} & 0.60 \\
            SCM\tnote{\textdagger} & 0.60 \\
            PROSO2\tnote{\textdagger} & 0.64 \\
            PaRSnIP\tnote{\textdagger} & 0.74 \\
            DeepSol\tnote{\textdagger} & 0.77 \\
        \bottomrule
    \end{tabular*}
    \begin{tablenotes}
            \item[\textdagger] Excerpted from \cite{khurana2018}
    \end{tablenotes}
    \end{threeparttable}
    \vspace{0.3cm}
\end{table}
\begin{table}[t!]
    \centering
 	\caption{Detailed localization prediction results}
  	\label{table:results_localization}
  	\footnotesize
    \begin{threeparttable}
    \begin{tabular*}{\columnwidth}{@{\extracolsep{\fill}}lc}
        \toprule
            Method & acc \\
        \midrule    
            \ptfm &  0.69  \\
            \plusb & 0.69  \\
            \plusl & \textbf{0.70}  \\
        \addlinespace[0.6ex] \cdashline{1-2} \addlinespace[0.6ex]
            P-ELMo &  0.54 \\
        \addlinespace[0.6ex] \cdashline{1-2} \addlinespace[0.6ex]
            SherLoc2\tnote{\textdagger} & 0.58 \\
            LocTree2\tnote{\textdagger} & 0.61 \\
            YLoc\tnote{\textdagger} & 0.61 \\
            iLoc-Euk\tnote{\textdagger} & 0.68 \\
            DeepLoc\tnote{\textdagger} & 0.78 \\
        \bottomrule
    \end{tabular*}
    \begin{tablenotes}
            \item[\textdagger] Excerpted from \cite{almagro2017}.
    \end{tablenotes}
    \end{threeparttable}
\end{table}
\begin{table}[t!]
    \centering
 	\caption{Detailed stability prediction results}
  	\label{table:results_stability}
  	\footnotesize
    \begin{threeparttable}
    \begin{tabular*}{\columnwidth}{@{\extracolsep{\fill}}lc}
        \toprule
            Method & $\rho$ \\
        \midrule    
            \ptfm & 0.76  \\
            \plusb & \textbf{0.77}  \\
            \plusl & \textbf{0.77}  \\
        \addlinespace[0.6ex] \cdashline{1-2} \addlinespace[0.6ex]
            P-ELMo\tnote{\textdagger} &  0.64 \\
            UniRep\tnote{\textdagger} &  0.73 \\
            TAPE-TFM\tnote{\textdagger} &  0.73 \\
            TAPE-RNN\tnote{\textdagger} &  0.69 \\
            TAPE-ResNet\tnote{\textdagger} &  0.73 \\
        \addlinespace[0.6ex] \cdashline{1-2} \addlinespace[0.6ex]
            RNN\textsubscript{BASE}    & 0.72 \\
            RNN\textsubscript{LARGE}   & 0.73 \\
        \bottomrule
    \end{tabular*}
    \begin{tablenotes}
            \item[\textdagger] Excerpted from TAPE \cite{rao2019}.
    \end{tablenotes}
    \end{threeparttable}
\end{table}
\begin{table}[t!]
    \centering
 	\caption{Detailed fluorescence prediction results}
  	\label{table:results_fluorescence}
  	\footnotesize
    \begin{threeparttable}
    \begin{tabular*}{\columnwidth}{@{\extracolsep{\fill}}lc}
        \toprule
            Method & $\rho$ \\
        \midrule    
            \ptfm & 0.63  \\
            \plusb & 0.67  \\
            \plusl & \textbf{0.68}  \\
        \addlinespace[0.6ex] \cdashline{1-2} \addlinespace[0.6ex]
            P-ELMo\tnote{\textdagger} &  0.33 \\
            UniRep\tnote{\textdagger} &  0.67 \\
            TAPE-TFM\tnote{\textdagger} &  \textbf{0.68} \\
            TAPE-RNN\tnote{\textdagger} &  0.67 \\
            TAPE-ResNet\tnote{\textdagger} &  0.21 \\
        \addlinespace[0.6ex] \cdashline{1-2} \addlinespace[0.6ex]
            RNN\textsubscript{BASE}    & 0.58 \\
            RNN\textsubscript{LARGE}   & 0.67 \\
        \bottomrule
    \end{tabular*}
    \begin{tablenotes}
            \item[\textdagger] Excerpted from TAPE \cite{rao2019}.
    \end{tablenotes}
    \end{threeparttable}
\end{table}

The detailed Solubility prediction results are presented in Table \ref{table:results_solubility}. We report the excerpted top-5 baseline results from \cite{khurana2018}. PaRSnIP \cite{rawi2018}, the second-best task-specific baseline model, used a gradient boosting classifier with over 8,000 1,2,3-mer amino acid frequency features, sequence-based features (\eg, length, molecular weight, and absolute charge), and structural features (\eg, secondary structures, a fraction of exposed residues, and hydrophobicity). DeepSol \cite{khurana2018}, the best task-specific baseline model, used a convolutional neural network consisting of convolution and max-pooling modules to learn sequence representations. It additionally adopted the sequence-based and structural features used in PaRSnIP to enhance the performance. 

\subsection{Localization} 
For the Localization task, we used the UniProt datasets \cite{apweiler2004}, which were pre-processed and provided by \cite{almagro2017}. The proteins were clustered with 30\% sequence identity. Then, each cluster of homologous proteins was split into 9,977 training, 1,108 development, and 2,773 test sequences. The subcellular locations are as follows: nucleus, cytoplasm, extracellular, mitochondrion, cell membrane, endoplasmic reticulum, plastid, Golgi apparatus, lysosome, and peroxisome. 

The detailed Localization prediction results are listed in Table \ref{table:results_localization}. We report the excerpted top-5 baseline results from \cite{almagro2017}. iLoc-Euk \cite{chou2011}, the second-best task-specific model, used a multi-label K-nearest neighbor classifier with pseudo-amino acid frequency features. Because iLoc-Euk predicted 22 locations, these were mapped onto our 10 locations. DeepLoc \cite{almagro2017}, the best task-specific model, used a convolutional neural network to learn motif information and a recurrent neural network to learn sequential dependencies of the motifs. It also adopted sequence-based evolutionary features through the combination of BLOSUM62 encoding \cite{eddy2004} and homology protein profiles from the Swiss-Prot database \cite{bairoch2000}. 

\subsection{Stability}
For the Stability task, we utilized the datasets from \cite{rocklin2017}, which were pre-processed by \cite{rao2019}. The dataset was split into 53,679 training, 2,447 development, and 12,839 test sequences. The test set contained one-Hamming distance neighbors of the top candidates from the training set. This allowed us to evaluate the model's ability to localize information from a broad sampling of relevant sequences.

The detailed Stability prediction results are listed in Table~\ref{table:results_stability}. As the data splits for this task were created by \cite{rao2019}, no clear task-specific SOTA existed. Instead, we present the results obtained using RNN\textsubscript{BASE} and RNN\textsubscript{LARGE} models without pre-training. 

\subsection{Fluorescence} 
For the Fluorescence task, we used the datasets from \cite{sarkisyan2016}, which were pre-processed by \cite{rao2019}. The dataset was split into 21,446 training, 5,362 development, and 27,217 test sequences. The training dataset contained three-Hamming distance mutations, whereas the test dataset contained more than four mutations. This allowed us to evaluate the model's ability to generalize to unseen mutation combinations.

The detailed Fluorescence prediction results are presented in Table \ref{table:results_fluorescence}. As this task were created by \cite{rao2019}, no clear task-specific SOTA exists. Instead, we present the results obtained using RNN\textsubscript{BASE} and RNN\textsubscript{LARGE} models without pre-training. 

\begin{table*}[t]
 	\caption{Ablation studies on Homology and SecStr tasks}
  	\label{table:result_ablation}
  	\footnotesize
    \begin{threeparttable}
    \begin{tabular*}{\textwidth}{l@{\extracolsep{\fill}}cccccccccc}
        \toprule
            & & & \multicolumn{6}{c}{Homology} & \multicolumn{2}{c}{SecStr} \\
        \cmidrule(){4-9} \cmidrule(){10-11}
            Method & $\lambda_{\mathrm{PT}}$ & $\lambda_{\mathrm{FT}}$ & acc & $\rho$ & Class & Fold & Superfamily & Family & acc8 & acc3 \\  
        \midrule    
            \plusb\ & 0.7 & 0.3 & \textbf{0.96} & \textbf{0.70} & \textbf{0.95} & \textbf{0.91} & \textbf{0.96} & \textbf{0.72} & \textbf{0.66} & \textbf{0.78} \\
        \addlinespace[0.6ex] \cdashline{1-11} \addlinespace[0.6ex]    
            RNN\textsubscript{BASE} & - &  0.3  & 0.93 & 0.67 & 0.88 & 0.81 & 0.92 & 0.68 & 0.61 & 0.73 \\
            (PT-A)	& 0.0 &  0.3  & 0.94 & 0.68 & 0.91 & 0.85 & 0.93 & 0.70 & 0.62 & 0.74 \\
            (PT-B)	& 0.5 &  0.3  & \textbf{0.96} & 0.69 & \textbf{0.95} & \textbf{0.91} & 0.95 & 0.70 & 0.66 & 0.77 \\
            (PT-C)	& 1.0 &  0.3  & \textbf{0.96} & 0.69 & 0.93 & 0.89 & 0.95 & 0.70 & 0.65 & 0.77 \\
        \addlinespace[0.6ex] \cdashline{1-11} \addlinespace[0.6ex]
            (FT-A)  &  0.7  & 0.0 & 0.94 & 0.68 & 0.91 & 0.85 & 0.93 & 0.70 & 0.65 & 0.77 \\
            (FT-B)  &  0.7  & 0.5 & \textbf{0.96} & 0.69 & \textbf{0.95} & \textbf{0.91} & 0.95 & 0.70 & \textbf{0.66} & \textbf{0.78} \\
        \bottomrule
    \end{tabular*}
    \begin{tablenotes}
        {\it Note:} We use the development sets for the ablation studies.
    \end{tablenotes}
    \end{threeparttable}
\end{table*}
\begin{table}[t!]
    \centering
 	\caption{Detailed transmembrane prediction results}
  	\label{table:results_transmembrane}
  	\footnotesize
    \begin{threeparttable}
    \begin{tabular*}{\columnwidth}{@{\extracolsep{\fill}}lc}
        \toprule
            Method & acc \\
        \midrule    
            \ptfm &  0.82  \\
            \plusb & \textbf{0.89}  \\
            \plusl & \textbf{0.89}  \\
        \addlinespace[0.6ex] \cdashline{1-2} \addlinespace[0.6ex]
            P-ELMo &  0.78 \\
        \addlinespace[0.6ex] \cdashline{1-2} \addlinespace[0.6ex]
            MEMSAT-SVM\tnote{\textdagger} & 0.67 \\
            Philius\tnote{\textdagger} & 0.70 \\
            SPOCTOPUS\tnote{\textdagger} & 0.71 \\
            TOPCONS\tnote{\textdagger} & 0.80 \\
        \bottomrule
    \end{tabular*}
    \begin{tablenotes}
            \item[\textdagger] Excerpted from P-ELMo \cite{bepler2019}.
    \end{tablenotes}
    \end{threeparttable}
\end{table}
\begin{figure}[t]
    \centering
	\includegraphics[width=\columnwidth]{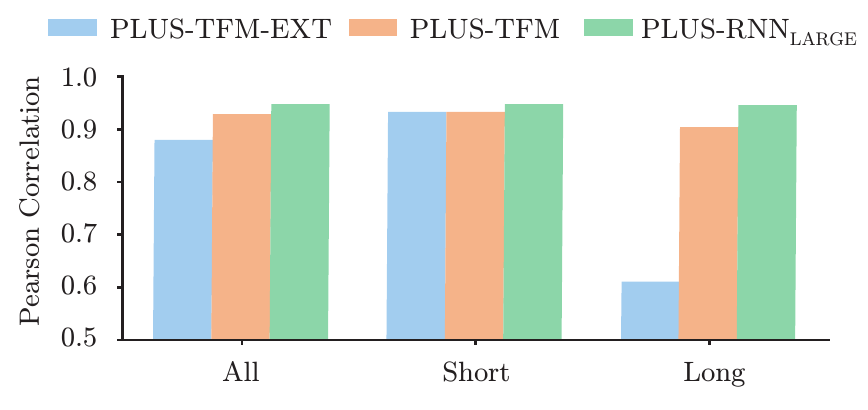}
   	\caption{Homology prediction results for different lengths.}
    \label{fig:homology_length}
\end{figure}

\subsection{Secondary structure (SecStr)} 
For the SecStr task, we used the training and development datasets from the PDB \cite{berman2003}, which were pre-processed and provided by \cite{rao2019}. Any sequences with more than 25\% sequence identity within the datasets were removed to avoid any bias due to homologous sequences. The training and development datasets contained 8,678 and 2,170 protein sequences, respectively. We used three test datasets: CB513 with 513 sequences \cite{cuff1999}, CASP12 with 21 sequences \cite{abriata2018}, and TS115 with 115 sequences \cite{yang2018}.

The detailed SecStr prediction results are presented in Table 3. We report excerpted results from \cite{rao2019} and \cite{klausen2019}. RaptorX \cite{wang2016} used a convolutional neural network to capture structural information and conditional neural fields to model secondary structure correlations among adjacent amino acids. It adopted the position-specific scoring matrix evolutionary features generated by searching the UniProt database \cite{apweiler2004} with PSI-BLAST \cite{altschul1997}. NetSurfP-2.0 \cite{klausen2019} used a convolutional neural network to learn motif information and a biRNN to learn their sequential dependencies. It adopted HMM evolutionary features from HH-bits \cite{remmert2012} including amino acid profiles, state transition probabilities, and local alignment diversities.

\subsection{Transmembrane} 
For the Transmembrane task, we used the TOPCONS datasets \cite{tsirigos2015}, which were pre-processed and provided by \cite{bepler2019}. The dataset was split into 228 training, 29 development, and 29 test sequences. The goal was to classify each amino acid into one of the following: the membrane, inside or outside of the membrane domains. 

The detailed Transmembrane prediction results are presented in Table \ref{table:results_transmembrane}. We report the excerpted top-5 baseline results from \cite{bepler2019}. Although this is an amino-acid-level prediction task, the evaluation was performed at the protein level, following the guidelines from TOPCONS. The predictions were judged correct if the protein had the same number of predicted and true transmembrane regions, and the predicted and true regions overlapped by at least five amino acids. TOPCONS \cite{tsirigos2015}, the best task-specific baseline model, is a meta-predictor that combines the predictions from five different predictors into a topology profile based on a dynamic programming algorithm.

\section{Ablation Studies}

\label{sec:appendix_ablation}

Here, we show results from ablation studies on the Homology and SecStr tasks to better understand the strengths and aspects of the \plus\ framework. We used \plusb\ as the baseline model unless explicitly stated otherwise. Note that we used the development sets for the ablation studies. 

\subsection{Pre-training and fine-tuning of \prnn}
We explored the effect of using different $\lambda_{\mathrm{PT}}$ values for controlling the relative importance of the MLM and SFP pre-training tasks (Table \ref{table:result_ablation}). The results indicated that the pre-trained models with different $\lambda_{\mathrm{PT}}$ values (\plusb, PT-A, PT-B, PT-C) always outperformed the RNN\textsubscript{BASE} model trained from the scratch. Both pre-training tasks consistently improve the prediction performance at all structural levels. Of the two pre-training tasks, removing MLM negatively affects the prediction performance more than removing the SFP. This coincides with the expected result, according to which, the MLM task would play the primary role, and the SFP task would complement MLM by encouraging the models to compare pairwise protein representations.

During the fine-tuning, we simultaneously trained a model for the MLM task as well as the downstream task. Moreover, we explored the effect of using different $\lambda_{\mathrm{FT}}$ values for controlling their relative importance (Table \ref{table:result_ablation}). The results showed that the models simultaneously fine-tuned with the MLM task loss (\plusb\ and FT-B) consistently outperformed the (FT-A) model fine-tuned only with the task-specific loss. Based on this, we infer that the MLM task serves as a form of regularization and improves the generalization performance of the models. 

\subsection{Comparison of \prnn\ and \ptfm}
We compared the Homology prediction performances of \ptfm\ and \plusl\ for protein pairs of different lengths (Figure \ref{fig:homology_length}). Because \ptfm\ was pre-trained using protein pairs shorter than 512 amino acids, we denote {\it Long} for protein pairs longer than 512 amino acids and {\it Short} otherwise. Next, we evaluated \ptfm\ for the {\it Long} protein pairs in the following two ways. First, we simply used the protein pairs as they were. Second, we truncated them to 512 amino acids. The former is denoted as PLUS-TFM-EXT (as in extended), and the latter is denoted as \ptfm. 

\plusl\ consistently provided competitive performance regardless of the protein length. In contrast, PLUS-TFM-EXT deteriorated for the {\it Long} protein pairs, whereas \ptfm\ exhibited a relatively less performance degradation. The results presented the limitations of TFM models using the limited context size of 512 amino acids. Although the number of {\it Long} protein pairs in the Homology development dataset was relatively small (13.4\%), complex proteins that are found in nature make the ability to analyze long protein sequences indispensable. Moreover, because this is due to the computational burden of TFM scaling quadratically with the input length, we predict that the recently proposed adaptive attention span approach \cite{sukhbaatar2019} may be able to help improve \ptfm.

\bibliographystyle{IEEEtran}
\bibliography{references}

\clearpage
\begin{IEEEbiography}[{\includegraphics[width=1in,height=1.25in,clip,keepaspectratio]{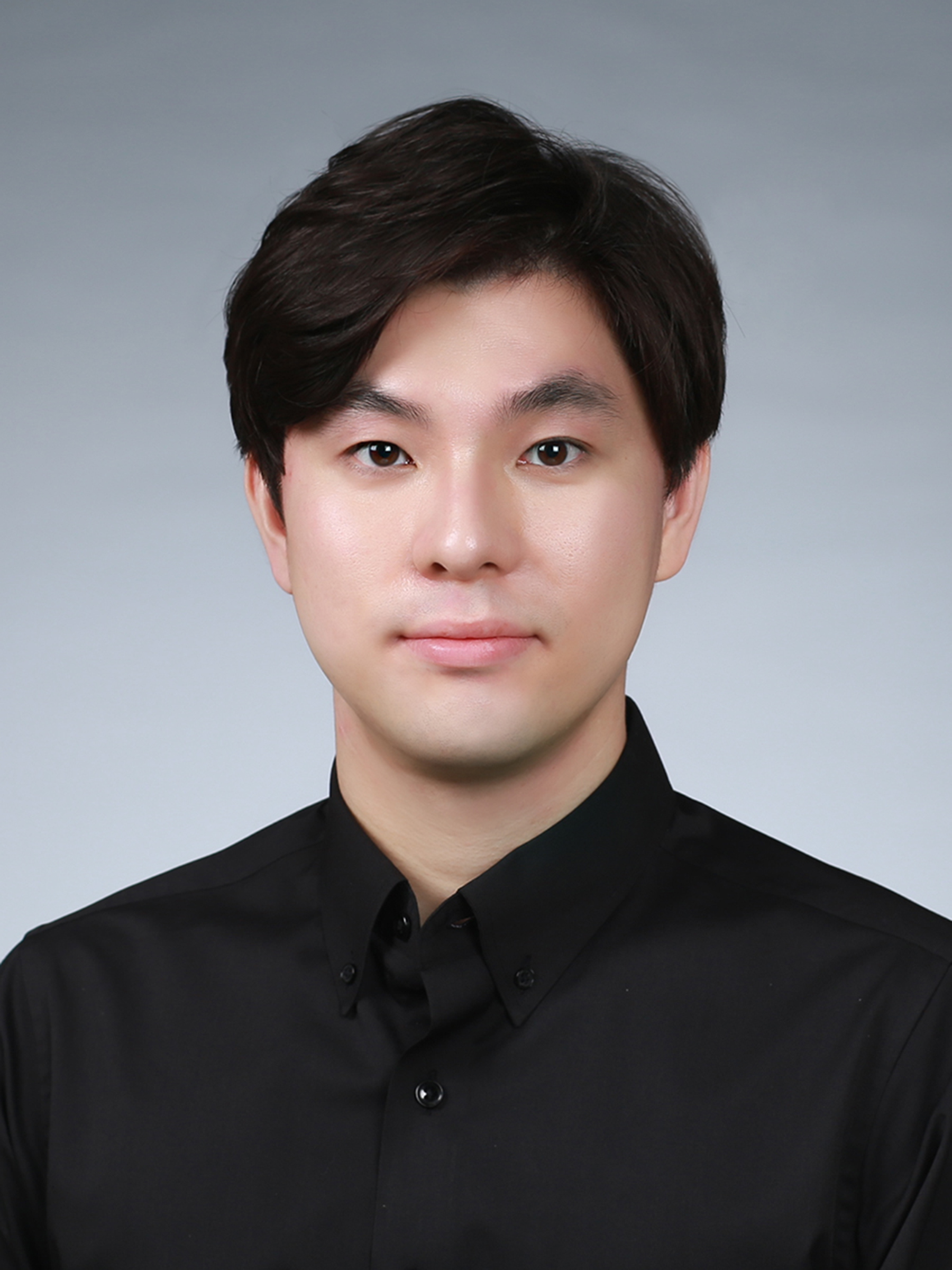}}]{Seonwoo Min} received the B.S. and Ph.D. degrees in electrical and computer engineering from Seoul National University, Seoul, Korea, in 2015 and 2021, respectively. He is currently a research scientist at Fundamental Research Laboratory, LG AI research. His current research interests include machine learning and bioinformatics. He was a recipient of the Global Ph.D Fellowship in 2016, the Microsoft Research Asia Fellowship Nomination Award in 2018, the Korea National Excellent Researcher Award in 2020, and many other prestigious awards. 
\end{IEEEbiography}

\begin{IEEEbiography}[{\includegraphics[width=1in,height=1.25in,clip,keepaspectratio]{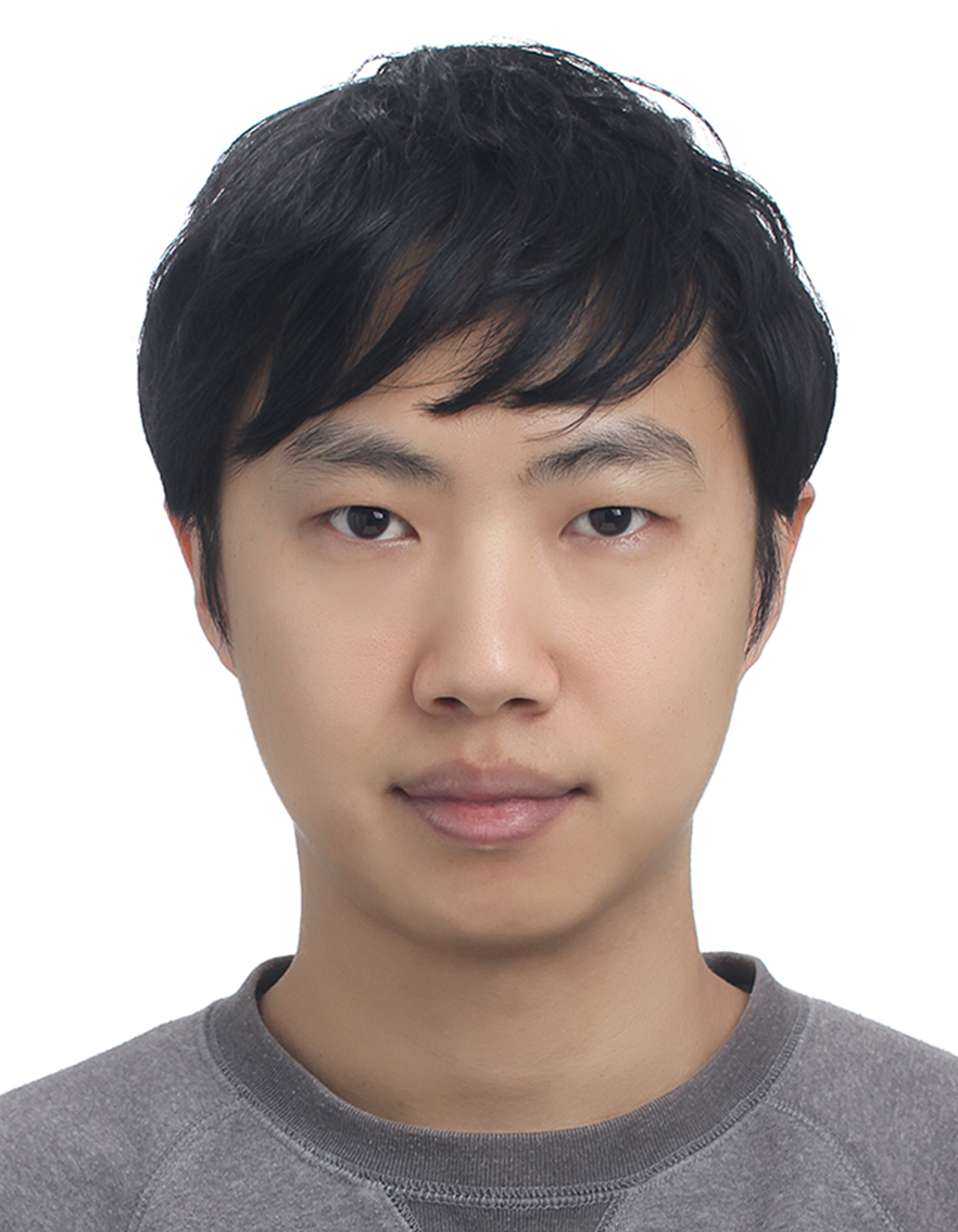}}]{Seunghyun Park} received the B.S. degree in Electrical Engineering from Korea University, Korea, in 2009. He received the Ph.D. degree in Electrical and Computer Engineering from Korea University, Korea, in 2018. Dr. Park also was a researcher at the Electrical and Computer Engineering, Seoul National University, Korea, from 2011 to 2018. Currently he is a research engineer at Clova AI, NAVER Corp., Korea. His research interests are machine learning, natural language processing, computational methods in statistics and medical informatics.
\end{IEEEbiography}

\begin{IEEEbiography}[{\includegraphics[width=1in,height=1.25in,clip,keepaspectratio]{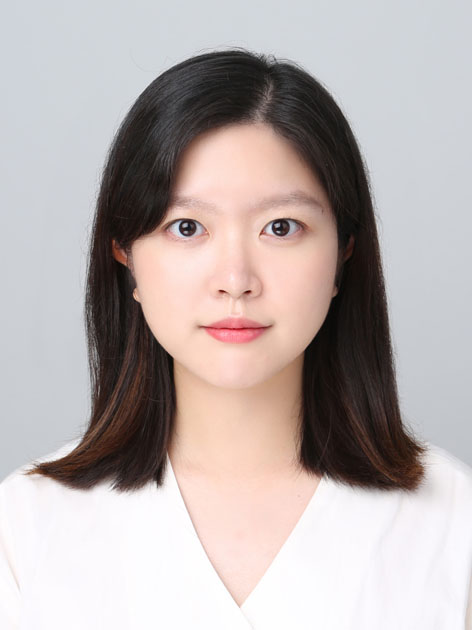}}]{Siwon Kim} received the B.S. degree in electrical and computer engineering from Seoul National University in Seoul, Korea in 2018. Currently, she is pursuing an integrated M.S./Ph.D. degree  in  electrical  and  computer  engineering  at Seoul National University. Her research interests include deep learning, explainable AI and biomedical applications.
\end{IEEEbiography}

\begin{IEEEbiography}[{\includegraphics[width=1in,height=1.25in,clip,keepaspectratio]{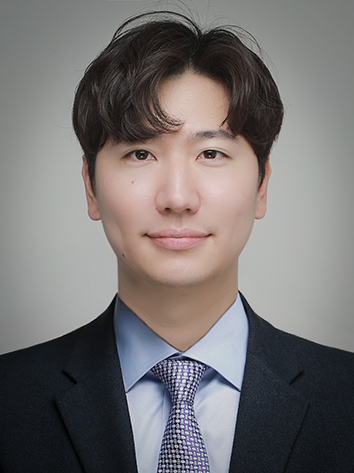}}]{Hyun-Soo Choi} received the B.S. degree in computer and communication engineering for the first major and in brain and cognitive science for the second major from Korea University, in 2013. He also received the integrated M.S./Ph.D. degree in electrical and computer engineering from Seoul National University, South Korea in 2020. From 2020 to 2021, he was a senior researcher in Vision AI Labs of SK Telecom. He is currently an Assistant Professor with the Department of Computer Science and Engineering, Kangwon National University, South Korea.
\end{IEEEbiography}

\begin{IEEEbiography}[{\includegraphics[width=1in,height=1.25in,clip,keepaspectratio]{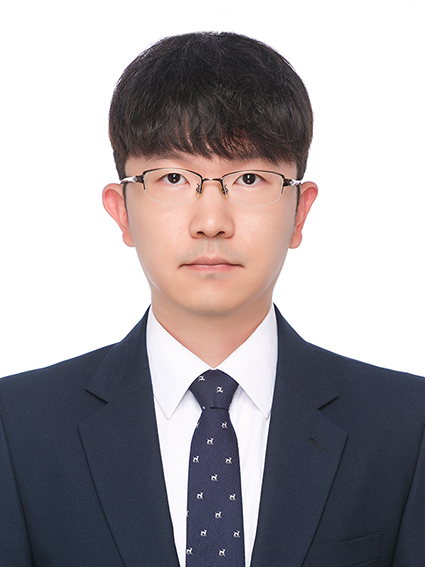}}]{Byunghan Lee}
(Member, IEEE) received the B.S. degree in electrical engineering from Korea University, South Korea, in 2011, and Ph.D. degree in electrical and computer engineering from Seoul National University, South Korea, in 2018. He is currently an Assistant Professor with the Department of Electronic and IT Media Engineering, Seoul National University of Science and Technology. His research interests include machine learning, artificial intelligence, and their biomedical applications.
\end{IEEEbiography}

\begin{IEEEbiography}[{\includegraphics[width=1in,height=1.25in,clip,keepaspectratio]{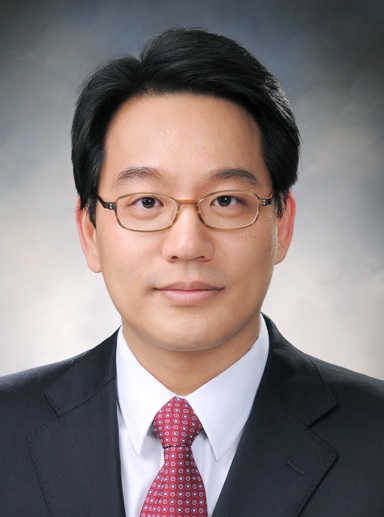}}]{Sungroh Yoon}
(Senior Member, IEEE) received the B.S. degree in electrical engineering from Seoul National University, South Korea, in 1996, and the M.S. and Ph.D. degrees in electrical engineering from Stanford University, Stanford, CA, USA, in 2002 and 2006, respectively. From 2006 to 2007, he was with Intel Corporation, Santa Clara. From 2007 to 2012, he was an Assistant Professor with the School of Electrical Engineering, Korea University. From 2016 to 2017, he was a Visiting Scholar with the Department of Neurology and Neurological Sciences, Stanford University. He held research positions at Stanford University and Synopsys, Inc., Mountain View. He is currently a Professor with the Department of Electrical and Computer Engineering, Seoul National University. His current research interests include machine learning and artificial intelligence. He was a recipient of the IEEE Young IT Engineer Award in 2013, the SBS Foundation Award in 2016, the IMIA Best Paper Award in 2017, the SNU Education Award in 2018, the IBM Faculty Award in 2018, the Korean Government Researcher of the Month Award 2018, the BRIC Best Research of the Year in 2018, the Shin-Yang Engineering Research Reward in 2019, the Microsoft Collaborative Research Grant in 2017 and 2020, and many other prestigious awards. Since February 2020, he has been serving as the Chairperson (Minister) for the Presidential Committee on the Fourth Industrial Revolution established by the Korean Government.
\end{IEEEbiography}

\EOD
\end{document}